\documentclass[journal=jacsat,manuscript=article]{achemso}

\usepackage[version=3]{mhchem} 

\usepackage{graphicx}
\usepackage{dcolumn}
\usepackage{bm}
\usepackage[utf8]{inputenc}
\usepackage[T1]{fontenc}
\usepackage{mathptmx}
\usepackage{amsmath,amssymb}
\usepackage{bm}
\usepackage{cleveref}
\usepackage{booktabs}
\usepackage{notes2bib}
\usepackage{braket}
\usepackage{multirow}
\usepackage{xcolor}
\usepackage[english]{babel}
\usepackage{verbatim}

\makeatletter
\def\@email#1#2{%
 \endgroup
 \patchcmd{\titleblock@produce}
 {\frontmatter@RRAPformat}
 {\frontmatter@RRAPformat{\produce@RRAP{*#1\href{mailto:#2}{#2}}}\frontmatter@RRAPformat}
 {}{}
}%

\DeclareUnicodeCharacter{2212}{-}
\DeclareUnicodeCharacter{2061}{}

\def\h2o{\mathrm{H}_2\mathrm{O}}

\author{Dario Baum}
\affiliation{Department of Chemistry and Pharmaceutical Sciences, Vrije Universiteit Amsterdam, De Boelelaan 1108, 1081 HZ Amsterdam, The Netherlands}%

\author{Lucas Visscher}
\affiliation{Department of Chemistry and Pharmaceutical Sciences, Vrije Universiteit Amsterdam, De Boelelaan 1108, 1081 HZ Amsterdam, The Netherlands}%

\author{Arno Förster}
\email{a.t.l.foerster@vu.nl}
\affiliation{Department of Chemistry and Pharmaceutical Sciences, Vrije Universiteit Amsterdam, De Boelelaan 1108, 1081 HZ Amsterdam, The Netherlands}%

\title{Predicting complete basis set limit quasiparticle energies from triple-$\zeta$ calculations}


\begin{document}

\begin{abstract}
We present a simple linear model to estimate the basis set incompleteness errors (BSIE) of (vertex-corrected) $GW$ QP energies based on the kinetic energy of the corresponding orbital only. We parametrise the model for $G_0W_0$, quasi-particle self-consistent $GW$ (qs$GW$), and vertex-corrected ($\Sigma^{BSE}@L^{BSE}$) QP energies on a large set of molecules containing 10 different elements for which we calculate complete basis set (CBS) limit extrapolated reference values with correlation-consistent basis sets ranging from triple- to hextuple-$\zeta$ (TZ/6Z). Based on these accurate reference values, we obtain model parameters for Gaussian-type and Slater-type orbital (GTO/STO) basis sets which allow for the extrapolation of QP energies calculated with TZ basis sets to the CBS limit with errors of 20 to 30 meV. Analysing extrapolation errors, we show the commonly used extrapolation method which assumes an inverse linear dependence of the BSIE on the inverse number of basis functions to be valid, but to produce larger errors, even when a quadruple-$\zeta$ calculation is used in the extrapolation. 
\end{abstract}

The $GW$ approximation (GWA) to the electronic self-energy\cite{ Hedin1965, martin2016} is widely used to calculate molecular quasiparticle (QP) energies\cite{Reining2018, Blase2018, photoemission_2, Marie2024b} which can be measured in (inverse) photoemission experiments and correspond to ionization potentials (IP) and electron aﬃnities (EA). At a much lower computational cost than that of more advanced wave function-based methods, molecular QP energies calculated with the best available $GW$ approaches show average deviations to highly accurate reference values of about 100 meV on average for first IPs\cite{Knight2016, Caruso2016, Bruneval2021a, McKeon2022, Bruneval2024, Forster2022, Forster2025} and of 200-300 meV for core IPs.\cite{Golze2020, Li2022a} 

The slow convergence to the complete basis set (CBS) limit with respect to the size of the single-particle basis,\cite{Bruneval2012, Bruneval2013, Bruneval2016a, Hung2017, slow_basis_set_convergence_1} much slower than that of correlated wave function methods,\cite{Forster2025, Marie2024} is however a major limitation. While basis set incompleteness errors (BSIE) largely cancel for the calculations of energy differences and excitation energies,\cite{pasquier2025gaussian} individual QP energies are often of interest in practical applications,  
ranging from the determination of molecular redox potentials which can be linked to first IPs,\cite{Forster2021a, Belic2022} over the prediction of core ionization energies,\cite{VanSetten2018, Golze2018, Golze2020, Mejia-Rodriguez2022, Li2022a, Panades-Barrueta2023, Kehry2023,  Bruneval2024, Yoneyama2024} to modelling of band structures of molecule-metal interfaces.\cite{Thygesen2009, Liu2019, Adeniran2021, Zhang2023a}

Even though various schemes have been designed to overcome the slow basis set convergence of $GW$ QP energies,\cite{Nguyen2012, CBS_5, CBS_6} practical calculations typically rely on extrapolations to the complete basis set limit. The simple two-point extrapolation method of Helgaker et al.\cite{Bruneval2012, Knight2016, Helgaker_CBS_1, Helgaker_CBS_2} or exponential fits aside,\cite{Li2019b} most calculations assume a linear dependence on the inverse number of basis functions.\cite{VanSetten2013, vanSetten2015, VanSetten2018, Golze2018, Stuke2020, Golze2020, Li2022a, Forster2021, Forster2022, Belic2022, Fediai2023, slow_basis_set_convergence_1} These methods are appealing due to their simplicity and broad applicability, but as we will also demonstrate here, they require a calculation in at least a quadruple-$\zeta$ (QZ) basis set to provide reliable estimates of the CBS limit. Modern $GW$ implementations are capable of performing such calculations for molecules with over 100 atoms,\cite{Stuke2020} but they are time-consuming. Calculations with QZ or even larger basis sets are especially problematic for low-scaling implementations\cite{Wilhelm2018, Forster2020b, Forster2021a, Wilhelm2021, Duchemin2021a} which scale quartic with basis set size.\cite{pasquier2025gaussian} Consequently, $GW$ calculations for systems with several thousands of electrons performed with these implementations have so far been restricted to at most triple-$\zeta$ (TZ) basis sets.\cite{Duchemin2021a, Amblard2022, Forster2020b, Forster2021a, Forster2022c, Wilhelm2018, Wilhelm2021}

Recently, \citet{bruneval_extrapolation} developed a simple extrapolation technique for $G_0W_0$ QP energies that only requires a single calculation in a small basis set. Their method relies on the training of a linear model that correlates the BSIE of a given QP energy, calculated in a specific basis set, with orbital-specific descriptors. They demonstrated their model to predict $G_0W_0$@BHANDH QP energies at the CBS limit with errors below 50 meV based on a single aug-cc-pvDZ calculation. 

In this work, we we present a simplified version of their model, which relies on the orbital kinetic energy as the only descriptor. While \citet{bruneval_extrapolation} focussed on $G_0W_0$@BHANDH, we here not only present parametrizations for $G_0W_0$, but also for qs$GW$,\cite{Faleev2004, VanSchilfgaarde2006, Kotani2007} and the $\Sigma^{BSE}@L^{BSE}$ method. The latter is a vertex-corrected approach which solves a Bethe-Salpeter equation (BSE) with the static, first-order $GW$ kernel for the two-particle response function $L$ and adds the same kernel to $\Sigma$ directly.\cite{Forster2024} For all of these methods, we parametrize the linear model for several Dunning correlation-consistent (CC) Gaussian-type orbital (GTO) and Slater-type orbital (STO) basis sets. We focus here on occupied valence and semi-valence QP energies only, since they are arguably more important in practice than virtual ones.

Key to the parametrisation of the linear model is the availability of reliable reference data at the CBS limit for a diverse range of molecules. In particular, this reference data should include QP energies of molecules of varying size, including medium and large molecules, since the convergence rate of QP energies depends on molecular size.\cite{Forster2020b} For the same reasons that motivate building such a model in the first place, this data is very hard to obtain. To calculate their reference data, \citet{bruneval_extrapolation} used Dunning correlation-consistent (CC) basis sets\cite{Dunning1989} up to quintuple-$\zeta$ (5Z) and estimated the CBS limit via the exponential fit of \citet{Li2019b}. In our work, we provide improved reference values on which we parametrize our linear model: We first calculate QP energies with Dunning CC basis sets ranging from triple-$\zeta$ (TZ) to hextuple-$\zeta$ (6Z) basis sets for an extensive set of molecules containing 10 different elements. We then investigate the basis set convergence of each QP energy individually, determine an optimized fit function which reproduces the observed convergence behavior, and use this fit to extrapolate to the CBS limit. In this way, we obtain high-quality, CBS-limit extrapolated QP energies. We then use those as a reference to parametrize the linear model, which predicts QP energies at the CBS limit with errors of only 20-30 meV, based on a single calculation in a TZ basis set. We also show that these errors are smaller than the ones obtained with the common technique of extrapolating to the CBS limit based on TZ and QZ calculations, assuming a dependence on the inverse number of basis functions. 

As a by-product of this work, we also critically evaluate the commonly applied schemes for CBS-limit extrapolation as well as the basis set convergence of qs$GW$ calculations. As we will demonstrate and rationalize later on, qs$GW$ shows an erratic basis set convergence behavior, and therefore does not permit a reliable CBS extrapolation or functional fitting. For qs$GW$, we therefore take QP energies calculated with the 5Z basis set as reference data. 

We prepare two different datasets of molecules. One dataset consists of molecules sampled from the SPICE \cite{SPICE} and the VQM24 \cite{VQM24} datasets containing the elements H, B, C, N, O, F, Si, P, S, and Cl. This set is called “General” in the following. With this set of elements, we cover the chemical space in which $GW$ is predominantly applied. \cite{mol_benchmarks_5, GW_chem_space_1, mol_benchmarks_10, mol_benchmarks_12, GW_chem_space_2, mol_benchmarks_14, GW_chem_space_3, GW_chem_space_4, GW_chem_space_5, GW_chem_space_6}. The other dataset is restricted to contain only organic molecules with elements H, C, N, O or F sampled from the QM9 \cite{QM9_1, QM9_2} and PC9 \cite{PC9} datasets. This set is hereafter called “Organic”. We add that second dataset since we are especially interested in obtaining CBS results for that class of molecules for future work. For both sets, we perform calculations with basis sets ranging from triple-$\zeta$ (TZ) to 6Z quality to approach the CBS limit as closely as possible. We additionally provide fitted extrapolation parameters for the DZP,\cite{vanLenthe2003} TZP,\cite{vanLenthe2003} TZ2P,\cite{vanLenthe2003}  and TZ3P\cite{Forster2021} STO type basis sets. 

All calculations are performed with a development version of the Amsterdam modelling suite (AMS) using the BAND engine for all GTO and the ADF engine for all STO calculations.\cite{Baerends2025} Most calculations are performed using the analytical expression of the $GW$ approximation as described in detail in Ref.~\citenum{Bruneval2016a}. Only for qs$GW$ calculations with a 5Z and the STO basis sets, we use the low-scaling analytical continuation-based implementation\cite{Forster2020b, Forster2021a} with 32 imaginary time and frequency points and a 32-point Padé model.\cite{Vidberg1977} We use the imaginary time and frequency points as implemented in the GreenX library.\cite{Azizi2023, Azizi2024} To verify the independence of the convergence to the CBS limit from the starting point, we perform $G_0W_0$ calculations using the BhandhLYP ($G_0W_0$@BhandhLYP) and PBE0 ($G_0W_0$@PBE0) starting points, as implemented in libXC.\cite{Lehtola2018} The $\Sigma^{BSE}@L^{BSE}$ calculations were performed using a Hartree--Fock reference as described in Ref.~\citenum{Forster2024}. The effective single-particle Hamiltonian for the qs$GW$ calculations with GTOs is constructed according to Eq.~10 (mode A) in Ref.~\citenum{Kotani2007}. Further technical aspects are discussed below in combination with an analysis of our numerical results.

\citet{bruneval_extrapolation} demonstrated that simple molecular orbital features can be exploited to approximate the difference between QP energies based on Dunning-type basis sets and the CBS limit.\cite{bruneval_extrapolation} For the feature vectors, they took the elements H, C, N, O, F, P, S, Cl and the $s$- and $p$-type atomic orbitals into account. The feature vector $x_i \in \mathbb{R}^{17}$ of a given sample Kohn--Sham orbital $\phi_i$ consists of Mulliken projections of the orbital on each of the considered elements and atomic orbitals and the natural logarithm of the kinetic energy expectation value $t_i = \langle \phi_i | -\frac{\nabla^2}{2} | \phi_i \rangle$ of this orbital. Here, we propose a simplification. We use the natural logarithm of $t_i$ as the single feature value for each sample orbital. The loss function $J$ to be minimized with respect to the parameters $\alpha_0$ and $\alpha_1$ becomes
\begin{equation}
    J(\alpha, \lambda) = \sum_i \Bigl( \Delta \varepsilon^{\text{TZ}}_i - \alpha_0 - \alpha_1 \ln (t_i) \Bigr)^2 \, .
    \label{eq: loss function}
\end{equation}
The basis set incompleteness error $\Delta \varepsilon^{\text{X}}_i$ for basis set X to be predicted for the $i$-th sample is thereby given as $\Delta \varepsilon^{\text{X}}_i = \varepsilon^{\text{CBS}}_i - \varepsilon^{\text{X}}_i$.
Therefore, for a given molecule, we extrapolate a QP energy obtained in a given basis set to its CBS limit result according to
\begin{equation}
    \varepsilon_i^{\text{CBS}} = \varepsilon_i^{\text{X}} + \alpha_0 + \alpha_1 \ln(t_i) \, .
\end{equation}
Carrying out this linear regression requires reference QP energies as close to the CBS limit as possible. Since fully converging QP energies with respect to basis set size is only feasible for the simplest systems, we rely on basis set extrapolation to approximate the CBS limit. We assume that the BSIE can be described by fit a function of the form
\begin{equation}
    f(x) = a x^{-k} \, ,
    \label{eq: optimal fit function}
\end{equation}
where $x$ might denote the cardinality of the basis set, i.e. $x \in \{3, 4, 5, 6\}$ in our case. We will refer to this variant of extrapolation as 'X' in the following. A two-parameter fit (usually using $X =3,4$) with an exponent of $k=3$ is sometimes chosen for the extrapolation of molecular QP energies,\cite{Bruneval2012, Knight2016} but $k=2$ has been shown to be a better choice.\cite{Bruneval2016a, Hung2017} Another possibility is to utilize the number of basis functions N, which is itself a function of the atom types and the chosen basis set. We will refer to this variant as 'N', for which an exponent of $k=1$ is the most popular choice for the extrapolation of molecular QP energies.\cite{VanSetten2013, vanSetten2015, VanSetten2018, Golze2018, Stuke2020, Golze2020, Li2022a, Forster2021, Forster2022, Belic2022}

In the following, we will assess the reliability of the 'X' and 'N' variants for extrapolating QP energies. In this assessment, we do not assume a fixed exponential parameter but also consider the exponent $k$ as a fit parameter. We first discuss the one-shot methods, i.e. $G_0W_0$ and $\Sigma^{BSE}@L^{BSE}$. In the following, we only show results for $G_0W_0$@BHandHLYP, since $G_0W_0$@PBE0 behaves completely identically. This indicates that the model parameters we obtain for $G_0W_0$@BHandHLYP can be used in all $G_0W_0$ calculations, irrespective of the starting point. Throughout this study, the number of samples in fit (or training) and test sets is determined by how many samples pass the quality checks that we implemented to filter out spurious and unphysical results. Such results can arise due to issues in converging either the SCF with very large basis sets or the qs$GW$ calculations.

\vspace{12pt}
\begin{table}[hbt!]
\centering
    \caption{Number of fit samples, mean exponents, standard deviation and average deviation between energies based on optimal exponents for each sample and average exponents.}
    \label{tab: average exponents and deviations}
    \renewcommand{\arraystretch}{1.25}
    \setlength{\tabcolsep}{8pt}
    \begin{tabular}{c|c|cccc|cccc}
        \hline
        & & \multicolumn{4}{c|}{\underline{$G_0W_0$}} & \multicolumn{4}{c}{\underline{$\Sigma^{BSE}@L^{BSE}$}} \\
        Input & Molecules & $n_{\text{samples}}$ & $\langle k\rangle $ & $\sigma$ & $\langle\Delta E\rangle$ [eV] & $n_{\text{samples}}$ & $\langle k\rangle$ & $\sigma$ & $\langle\Delta E\rangle$ [eV] \\
        \hline
        \multirow{2}{1em}{X} & General & 160 & 1.6 & 0.3 & 0.0660 & 148 & 1.6 & 0.3 & 0.0375 \\
        & Organic & 317 & 2.0 & 0.2 & 0.0222 & 385 & 2.1 & 0.9 & 0.0219 \\
        \hline
        \multirow{2}{1em}{N} & General & 160 & 0.9 & 0.1 & 0.0286 & 147 & 0.9 & 0.1 & 0.0167 \\
        & Organic & 320 & 1.0 & 0.1 & 0.0169 & 393 & 1.0 & 0.1 & 0.0159 \\
    \end{tabular}
\end{table}
\vspace{12pt}

In Tab.~\ref{tab: average exponents and deviations} we list the number of fit samples, the mean exponent $\langle k \rangle$ and the respective standard deviation $\sigma$ for both molecule sets, based on the cardinality (“X”) and based on the number of basis functions (“N”) for the HOMO QP energy. We also list the average deviation $\langle\Delta E\rangle$ between the CBS energy that one would obtain using the mean exponent $\langle k \rangle$ and the CBS energy that one would obtain fitting the “optimal” exponent $k$ for each sample individually.

\begin{figure}[hbt!]
	\centering
	\includegraphics[width=\textwidth]{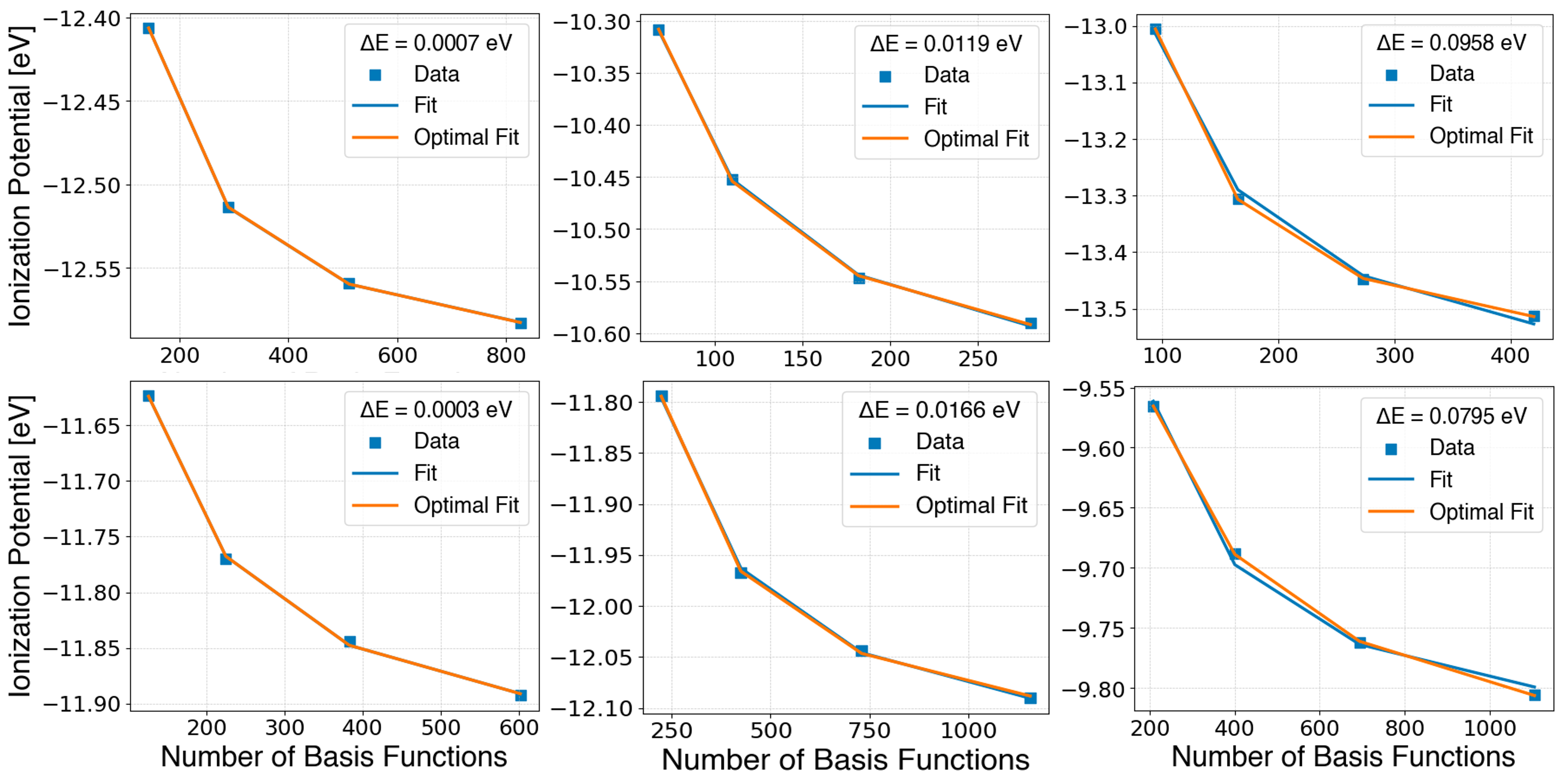}
	\caption{Basis set convergence with fits based on average exponent (“Fit”) and based on the individually optimal exponents (“Optimal Fit”) for $G_0W_0$@BHandHLYP (top row) and $\Sigma^{BSE}@L^{BSE}$ (bottom row) on the General set. Examples are shown for samples where, with respect to the average exponent, the fitted exponent exactly coincides (left column), deviates by one standard deviation (middle column) or deviates the most out of all samples in the set (right column).}
	\label{fig: best and worst performers}
\end{figure}

There are two key observations. First, the commonly applied exponent of $k=3$ based on basis set cardinality is inappropriate for $G_0W_0$ and $\Sigma^{BSE}@L^{BSE}$ in all cases tested by us. An exponent of $k \approx 2$ is more appropriate, confirming an earlier observation by Bruneval and co-workers.\cite{Bruneval2016a, Hung2017} Second, the fits based on cardinality X show higher variations compared to the fits based on the number of basis functions N. For the latter, both molecule sets seem to agree on an exponent of $k=1$ for $G_0W_0$ and $\Sigma^{BSE}@L^{BSE}$. Irrespective of whether 'N' or 'X' is chosen, the extrapolation errors made using the respective mean exponents are fairly small, i.e. in the range of a few ten meV. In Fig.~\ref{fig: best and worst performers}, we present for both $GW$ methods from Tab.~\ref{tab: average exponents and deviations} examples of the basis set convergence of the samples where the average exponent is spot on (left column), where the optimal exponent deviates from the average one by one standard deviation (middle column) and there worst respective outlier (right column). This demonstrates that the deviations in energy are still acceptable even for the most severe outliers. To obtain accurate reference CBS energies, we fit individual and therefore optimal exponents for each fit and test sample. In other words, for each sample, we fit \cref{eq: optimal fit function} to the respective TZ, QZ, 5Z and 6Z calculations for each molecule with its individual exponent.

\begin{figure}[hbt!]
	\centering
	\includegraphics[width=0.75\textwidth]{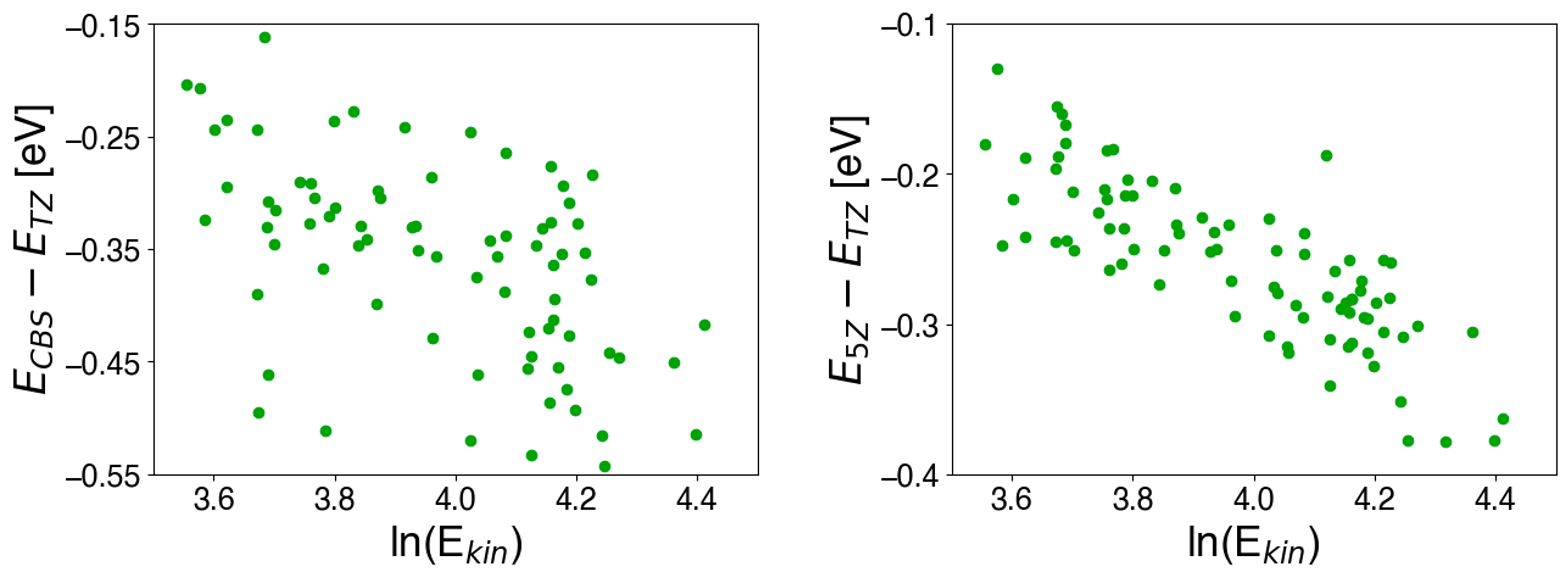}
	\caption{Linear correlation between the natural logarithm of the kinetic energy and the BSIE, based on the CBS limit (left) and based on a 5Z basis set (right).}
	\label{fig: recover qsGW linear correlation}
\end{figure}

Now we discuss our qs$GW$ results. Here we observe that the BSIE does not reliably follow the exponential decay usually assumed. Rather, the convergence sometimes is much faster than expected, rendering it ill-suited for the fit Eq.~\eqref{eq: optimal fit function}, sometimes much slower than expected, resembling more a linear than an higher polynomial decay, or instable altogether, with the BSIE increasing for QZ and then decreasing again for 5Z with respect to TZ. Even after filtering out instable convergence behavior, samples with much faster or much slower convergence lead to large variance in the fitted exponents (we observe a standard deviation of 0.6). Consequently, such broad spread of convergence rates reduces the quality of the linear correlation between the natural logarithm of the kinetic energy and the BSIE as shown in Fig. \ref{fig: recover qsGW linear correlation} on the left. Thus, extrapolation to the CBS limit comes with unacceptably large MAE of around 0.08\,eV. 

The issue can be linked to the intruder-state problem, well-known from multi-reference perturbation theories.\cite{hegarty1979calculation, evangelisti1987qualitative} 
In the notation of Ref.~\citenum{Marie2023a} and in the basis of QP states $p,q,r \dots$, the qs$GW$ hamiltonian used in this work is
\begin{equation}
\label{qsGWH}
\Sigma_{pq} = \frac{1}{2} 
\sum_{r \nu}
\left[
\frac{\Delta_{\nu, pr}}{\Delta^2_{\nu, pr} + \eta^2} + 
\frac{\Delta_{\nu, qr}}{\Delta^2_{\nu, qr} + \eta^2}
\right] W_{\nu, pr} W_{\nu, qr} \;,
\end{equation}
with $\Delta_{\nu, pr} = \epsilon_p - \epsilon_r - \text{sgn} (\epsilon_r-\epsilon_F)\Omega_{\nu}$, where $\epsilon$ denotes a QP energy, $\Omega_{\nu}$ is a neutral excitation energy in the random phase approximation (RPA), $W$ denotes a corresponding transition amplitude and $\epsilon_F$ is the Fermi-energy. $\eta$ is a positive real number.  
Eq.~\eqref{qsGWH} arises from downfolding an effective Hamiltonian in the $1h+1p + 2h1p + 2p1h$ space to the $1h+1p$ space (spanned by a single reference determinant).\cite{Bintrim2021, Tolle2023} When active space and environment are not distinguished by a clear energy-scale separation, some of the external configurations in the $2p1h/2h1p$ become energetically close to the reference determinant, leading to vanishing $\Delta$ in Eq.~\eqref{qsGWH}.\cite{Monino2022, Marie2023a} This makes the optimization landscape flat and non-smooth and can lead to convergence to different solutions both in the weakly and strongly correlated regime.\cite{Marie2023a, Ammar2024}  These issues in solving the  qs$GW$ equations manifest themselves in different ways when basis sets of varying sizes are used, thus making basis set extrapolation difficult. 

Numerical results supporting this observation as well as further numerical tests can be found in the SI. We furthermore investigated the effect of using different values of $\eta$ in Eq.~\eqref{qsGWH}. While using a rather large value of $\eta = 0.01$ Hartree as in Ref.~\citenum{Marie2023a} has a somewhat positive effect on the convergence, the erratic behaviour still persists. While using an even higher value of $\eta$ could potentially solve the problem, it would lead to a sizable artificial shift in the QP energies. For this reason, we do not investigate this approach to solve the issue.

Instead we use 5Z results as extrapolation references. As can be seen in Fig. \ref{fig: recover qsGW linear correlation} on the right, in that way the linear correlation between the logarithm of the kinetic energy and the BSIE (now based on 5Z instead of CBS) is much clearer. We choose 5Z because it is the highest basis set quality for which the corresponding calculations for fit data production are still somewhat feasible. Given the considerable computational cost of qs$GW$ calculations with a basis set of 5Z quality, such an extrapolation is still valuable. Although the convergence of the energies is overall quite erratic, the 5Z seems to give reliable results. We present evidence on that in the SI.

Using these reference energies obtained in this way (5Z for qs$GW$, CBS-limit extrapolated for $G_0W_0$ and $\Sigma^{BSE}$@$L^{BSE}$), we fit the extrapolation parameters $\alpha_0$ and $\alpha_1$ for both molecule sets by minimizing the least squares loss defined in Eq.~\ref{eq: loss function}. Again, note that for the CBS-extrapolation in the case of $G_0W_0$ and $\Sigma^{BSE}$@$L^{BSE}$ we do \textit{not} employ standard two-extrapolation but instead fit Eq.~\ref{eq: optimal fit function} to the respective TZ, QZ, 5Z, 6Z calculations for each molecule individually to obtain references as accurate as possible. We then evaluate the fitted parameters for our proposed extrapolation approach on separate test sets of molecules which have not been used for the fit and for which we prepared reference energies in the same way described before. The numbers of fit and test samples, the resulting parameters as well as the MAEs between the predicted CBS limit energies based on cc-pVTZ calculations and the reference energies on the test sets are given in the SI.

\begin{figure}[hbt!]
	\centering
	\includegraphics[width=0.9\textwidth]{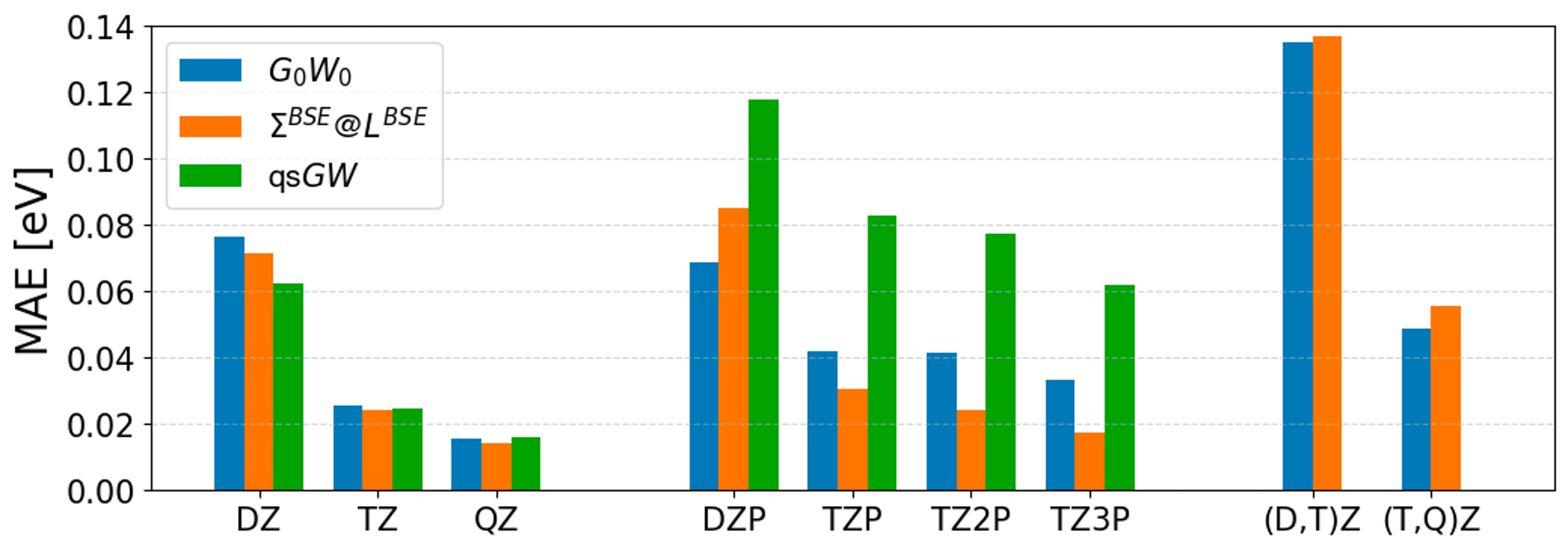}
	\caption{MAE of predicted BSIEs of QP energies on the Organic set using our extrapolation based on different GTO basis sets (left), STO basis sets (middle) and using standard two-point basis set extrapolation (right).}
	\label{fig: MAE vs basis set start point}
\end{figure}

We propose to use TZ calculations as starting points for approximating the CBS limit using our approach because it provides a good trade-off between affordability in computing the starting point and accuracy of the resulting extrapolation. To support this statement, we present the MAEs of the estimated BSIEs on the Organic test set when starting from a DZ, TZ and QZ calculation respectively in Fig. \ref{fig: MAE vs basis set start point} (on the left). Additionally, we show the MAEs of estimated BSIEs based on STO basis sets (in the middle), which we will discuss later. For a comparison, we also plot MAEs of BSIEs computed with standard two-point extrapolation using DZ plus TZ calculations and TZ plus QZ calculations respectively (on the right).

As expected, for all $GW$ methods, the error decreases with increasing basis set size. For all basis set sizes used here, the MAE for $\Sigma^{BSE}@L^{BSE}$ are the lowest which is due to the lower BSIE of $\Sigma^{BSE}@L^{BSE}$.\cite{Forster2025} Most importantly, based on a TZ calculation, we obtain extrapolation errors of on average less than 30\,meV. These errors are well below the general error bars of these methods compared to high-level wave function-based methods\cite{Knight2016, Caruso2016, Bruneval2021a, McKeon2022, Bruneval2024, Forster2022, Forster2025} and below the resolutions typically attained in photoemission experiments. With a TZ starting point for the extrapolation, we outperform the standard two-point CBS extrapolation based on a TZ and QZ basis set in our tests. Two-point extrapolation is inherently built to estimate the CBS limit, not a 5Z basis set. Therefore, we do not plot the result for standard two-point extrapolation for qs$GW$ in Fig. \ref{fig: MAE vs basis set start point} as it fails as expected and would thus give a misleading comparison.



\begin{figure}[H]
	\centering
	\includegraphics[width=\textwidth]{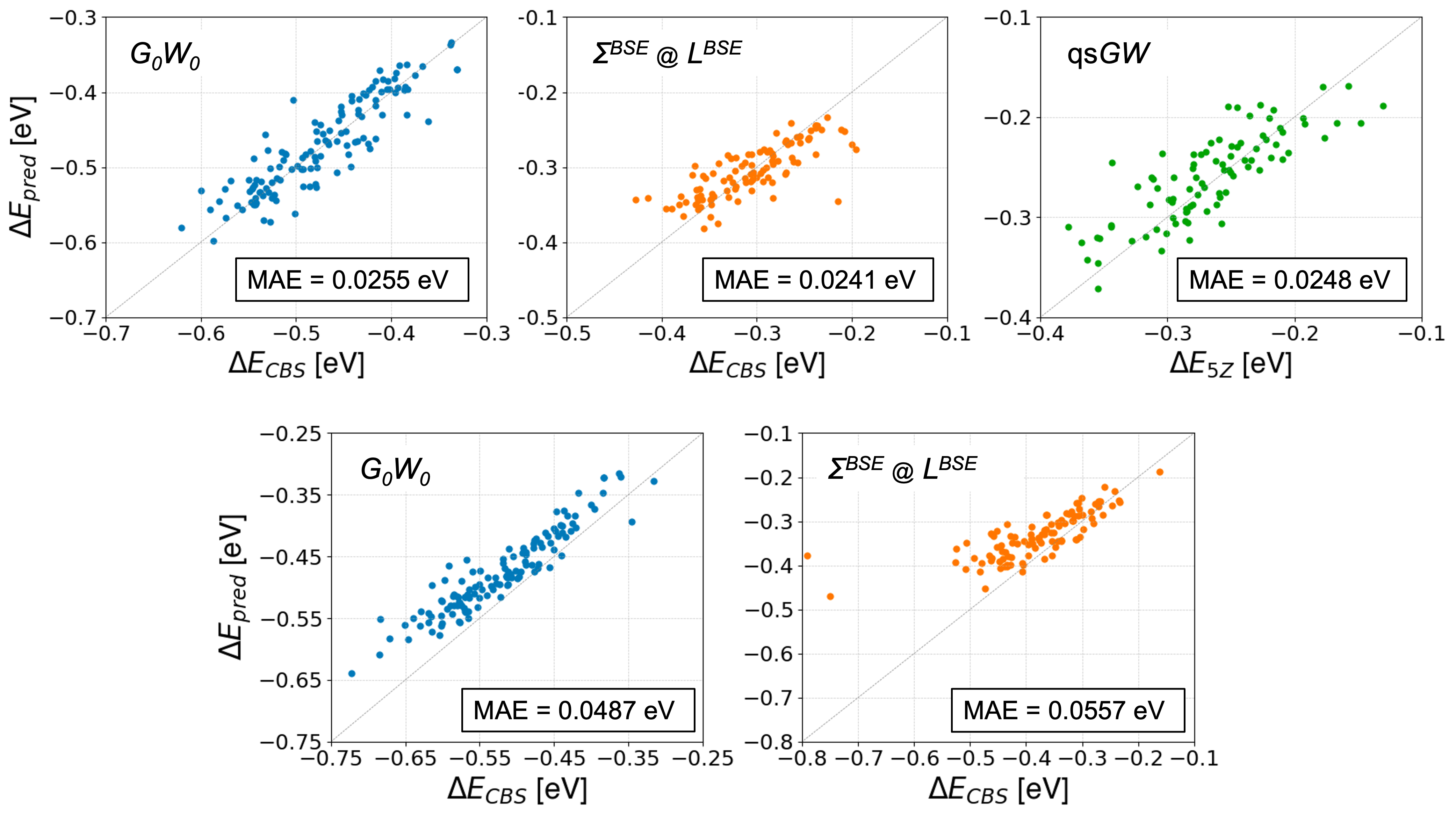}
	\caption{Predicted BSIEs for the Organic set, in the top row using our proposed extrapolation method based on TZ calculations, in the bottom row using standard two-point extrapolation with TZ and QZ calculations.}
	\label{fig: prediction accuracies}
\end{figure}

The combination of our extrapolation scheme with a TZ calculation represents an excellent trade-off between efficiency and accuracy. In the top row of fig.~\ref{fig: prediction accuracies}, we plot the predicted BSIEs based on TZ calculations against the reference BSIEs for $G_0W_0$, $\Sigma^{BSE}@L^{BSE}$, and qs$GW$ again exemplarily for the Organic set. This reiterates the excellent agreement between the predictions and the references and reveals that there are no severe outliers and only small deviations from the reference energies throughout. In the bottom row, Fig.~\ref{fig: prediction accuracies} shows the same for the standard two-point extrapolation, revealing a slightly larger spread of values, as well as some outliers in case of $\Sigma^{BSE}$@$L^{BSE}$.

Next, we show that we can also, in the same way, extrapolate results obtained using some STO basis set to the GTO results at the CBS limit. This approach is based on the assumption that calculations with different basis set types still converge to the same results in the limit of an infinite number of basis functions, which, to a good approximation, is known to be the case for IPs\cite{Maggio2017a, Govoni2018, Forster2021}. The parameters to extrapolate $G_0W_0$, $\Sigma^{BSE}@L^{BSE}$ and qs$GW$ calculations with different STO as implemented in ADF are listed in the SI. As already demonstrated in Fig.~\ref{fig: MAE vs basis set start point}, also with this approach, we outperform standard two-point extrapolation for $G_0W_0$ and $\Sigma^{BSE}@L^{BSE}$ for the three larger STO basis sets. For the smallest STO basis set (DZP), the MAEs are comparable to pure GTO extrapolation and still outperform standard extrapolation if also based on smaller basis sets (DZ and TZ). For qs$GW$, the errors are somewhat larger, although still acceptable, especially considering the provided speed up. For the TZ3P basis set, it is still roughly on par with standard extrapolation

The larger errors could be due to another caveat that we observed for qs$GW$: opposed to the other two methods, it seems that there is no linear correlation between the logarithm of the kinetic energy matrix elements and the BSIE. Specifically, we find only small values for the $\alpha_1$ parameter, indicating only minor dependence on the kinetic energy. 

\begin{figure}[hbt!]
	\centering
	\includegraphics[width=0.75\textwidth]{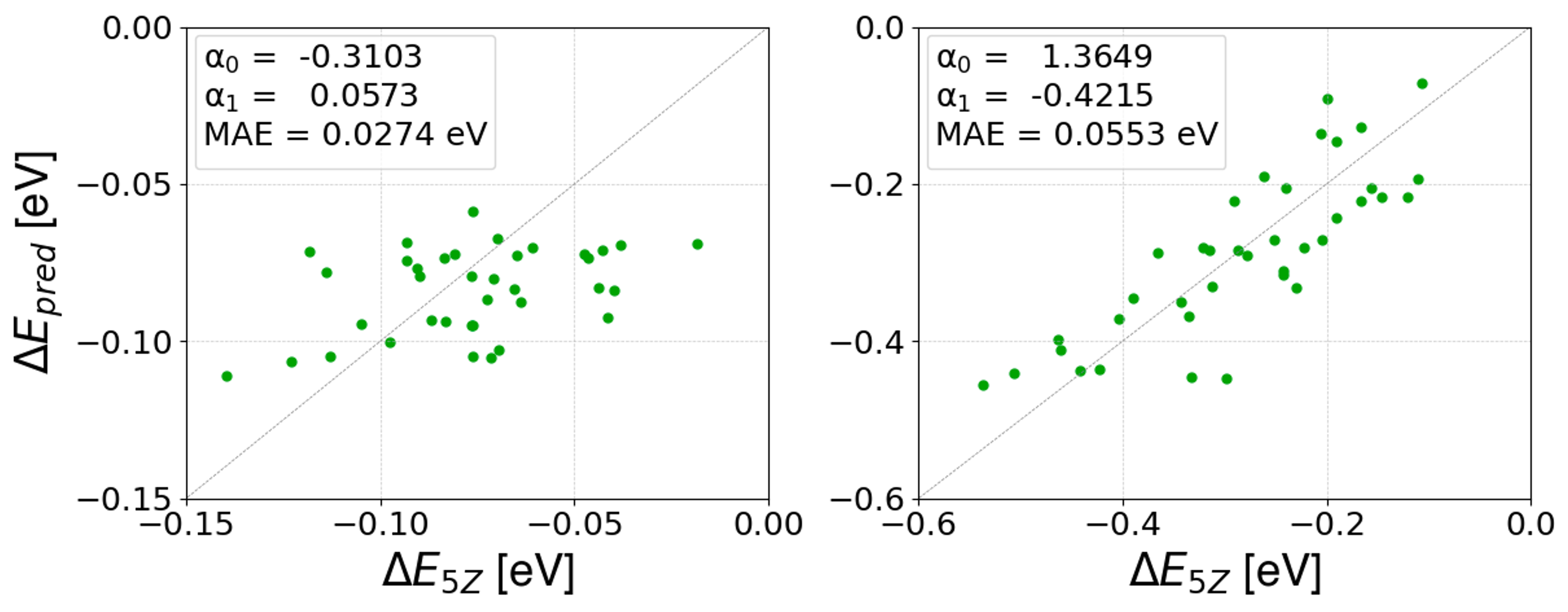}
	\caption{Predicted BSIE based on TZ3P calculations for the Organic set plotted against the corresponding reference BSIE based on converged QP energies (left) and based on first-iteration QP energies (right).}
	\label{fig: MAE vs basis set start point qsGW}
\end{figure}

Since extrapolating from the STO data to the GTO limit only fails for qs$GW$, we speculate that over the course of the self-consistent loops, differences between STO and GTO start to amplify, such that the linear correlation as described before eventually breaks down. To test this hypothesis, we extract the QP energies from the first qs$GW$ iterations of the STO calculations instead of the converged ones. We then perform our proposed linear regression with the BSIE being the difference between the (converged) 5Z GTO energies and the first-iteration QP energies in a given STO basis set. Fig.~\ref{fig: MAE vs basis set start point qsGW} shows the predicted BSIE plotted against the ground truth targets, based on converged STO energies on the left and based on first-iteration STO energies on the right with the fitted extrapolation parameters, respectively. The non-self-consistently obtained STO energies show a good linear correlation to the self-consistently obtained GTO-5Z results, and the extrapolation parameters qualitatively agree with other parameters fitted in this work. The range of deviations is wider in the right plot, because the differences to the extrapolation target due to non-self-consistency and BSIE add up. All in all, this finding allows for an even more efficient basis set extrapolation scheme: instead of running a full STO qs$GW$ calculation and then extrapolating it to a 5Z GTO result, one needs to run only one iteration of such a calculation and can immediately extrapolate that to a \textit{converged} 5Z GTO result. Note that the fitted extrapolation parameters $\alpha_0$ and $\alpha_1$ for qs$GW$ in the SI for STO basis sets are based on this approach of using the first qs$GW$ iteration.

To summarise, we parametrize a conceptually simple model which contains only two parameters and requires only the kinetic energy as input to extrapolate QP energies to the CBS limit. We demonstrate the accuracy and versatility of the approach and analyze the error dependence on the extrapolation starting point with respect to the basis set cardinality. As a result of this study, we provide parametrizations for $G_0W_0$, $\Sigma^{BSE}@L^{BSE}$ and qs$GW$ calculations based on Dunning-type GTO and STO basis sets, which are readily available in popular quantum chemistry programs. Using the reference values presented here, parametrizations for other basis sets can be readily obtained as well. We have also confirmed the expectation that the basis set convergence of $G_0W_0$ does not depend on the starting point. Therefore, the parameters we presented here for $G_0W_0$ can be used irrespective of the starting point. In this work we have restricted ourselves only to valence IPs since they are relevant in many applications. However, the same scheme could work even better for core IPs since core electrons have a higher kinetic energy density than the more diffuse valence orbitals.\cite{cohen1961cancellation}

To obtain reliable reference QP energies for fitting the extrapolation parameters, we critically evaluated the basis set convergence of (vertex-corrected) $GW$ calculations. The extrapolation method of Helgaker and coworkers, which assumes an inverse cubic dependence of the BSIE on basis set cardinality, substantially overestimates the basis set convergence. Instead, choosing an exponent closer to $k=2$ is more appropriate.\cite{Bruneval2016a, Hung2017} The commonly performed extrapolation of the QP energies assuming an inverse linear dependence on the number of basis functions is justified for $G_0W_0$ and $\Sigma^{BSE}@L^{BSE}$. For qs$GW$, the situation is more nuanced. Here, the convergence to the CBS limit is more erratic. We do not think that this is a fundamental issue, but rather related to difficulties to converge the qs$GW$ equations to the same physical solution in different basis sets.

\begin{acknowledgement}
We acknowledge the use of supercomputer facilities at SURFsara sponsored by NWO Physical Sciences, with financial support from The Netherlands Organization for Scientific Research (NWO). AF acknowledges funding through a VENI grant from NWO under grant agreement VI.Veni.232.013.
\end{acknowledgement}


\bibliographystyle{plain}
\bibliography{references,all}

\end{document}